\documentclass[prd,onecolumn,nopacs,nofootinbib]{revtex4}

\usepackage{amsmath}
\usepackage{graphicx}
\usepackage{amsfonts}
\usepackage{latexsym}
\usepackage{bbold}
\usepackage{wasysym}
\usepackage{calligra}
\usepackage{float}
\usepackage{ulem}
\usepackage{inputenc}
\usepackage{xspace}
\usepackage{url}
\usepackage{epstopdf}
\usepackage{tikz}
\usepackage{amssymb}
\usepackage{enumitem}

\usepackage{amsmath}
\usepackage{amssymb}
\usepackage{graphicx}
\usepackage{amsfonts}
\usepackage{latexsym}
\usepackage{bbold}
\usepackage{wasysym}
\usepackage{calligra}
\usepackage{float}
\usepackage{ulem}
\usepackage{inputenc}
\usepackage{xspace}
\usepackage{url}
\usepackage{epstopdf}
\usepackage{tikz}
\usepackage{amsthm}
\usepackage{soul}

\newtheorem{cor}{Corollary}

\newtheorem{theorem}{Theorem}

\newcommand{\be}{\begin{equation}}
\newcommand{\ee}{\end{equation}}
\newcommand{\beq} {\begin{equation}}
\newcommand{\eeq} {\end{equation}}
\newcommand{\ba}{\begin{eqnarray}}
\newcommand{\ea}{\end{eqnarray}}

\newcommand{\td}{\tilde}
\newcommand{\mn}{\mu\nu}
\newcommand{\ab}{\alpha\beta}

\begin{document}

    \title{2-dimensional Metric-Affine Geometries and Dynamical Gravity }
	\author{Damianos Iosifidis}

\affiliation{Scuola Superiore Meridionale, Largo San Marcellino 10, 80138 Napoli, Italy},

\affiliation{INFN– Sezione di Napoli, Via Cintia, 80126 Napoli, Italy}

	\email{d.iosifidis@ssmeridionale.it}

	\date{\today}
	\begin{abstract}
We study non-Riemannian metric-affine geometries in 2-dimensions. We explicitly show that if the connection of the space is not metric compatible (i.e. the non-metricity is non-vanishing) then the underlying 2-dimensional manifold is not an Einstein manifold, in contrast to the Riemannian case. Furthermore we prove that in general the curvature in such spaces is specified completely by the Ricci tensor which, however, is not symmetric now. An enlarged symmetry of the scalar curvature under vectorial connection shifts is also found. We then apply our results to the study of 2-dimensional Metric-Affine Gravity Theories and discuss some special cases.
		
	\end{abstract}
	
	\maketitle
	
	\allowdisplaybreaks
	
	

\section{Introduction}

It is a widely well known mathematical result, and quite trivial to prove, that in 2 dimensions the full  Riemannian curvature is completely specified by the Riemannian Ricci scalar.
From a physical perspective, 
in this Riemannian geometry setting, the 2-dimensional Einstein-Hilbert action is topological; it has no local propagating degrees of freedom, and is therefore trivial, the Einstein tensor vanishes identically. This is so because in 2-dimensions, the Riemannian curvature, given its symmetries, has only one independent component and is therefore completely determined by the Ricci tensor through the well-known formula
\beq
\tilde{R}_{ijkl}=\frac{\tilde{R}}{2}(g_{ik}g_{jl}-g_{il}g_{jk})
\eeq
\beq
\td{R}_{ij}=\frac{\td{R}}{2}g_{ij} \label{RiemR}
\eeq
where the tildes denote Riemannian quantities computed with respect to the torsionless and metric-compatible Levi-Civita connection. In other words, what (\ref{RiemR}) tells us is that every 2-dimensional Riemannian manifold is an Einstein space.

In order to obtain non-trivial dynamics, an additional dilaton field has to  be included leading to the so-called 2-dimensional JT gravity \cite{Teitelboim1983,Jackiw1985}. \footnote{See  \cite{Grumiller:2002nm} for a comprehensive review.}
Alternatively, one may obtain non-trivial dynamics by enriching the underlying geometry by allowing for torsion and nonmetricity. This is the most well motivated geometric approach to 2-dimensional gravity. Quite surprisingly, not many studies have been done on the matter, with the notable exceptions \cite{Katanaev:1986qm,Mielke:1993nc,Obukhov:2003tz}.

In this note we provide some useful novel results related to the geometry of 2-dimensional metric-affine spaces. Given these results we then consider applications to gravity and show how one can have dynamical 2-dim gravity as a consequence of the generalized h=geometry.
In what follows the signature of the 2-dim metric will be $((-1)^s ,1)$ where $s=0$ correspond to a Eucledean and $s=1$ to Lorentzian manifold respectively. When we study gravity theories we will stick to the Lorentizan signature. The
Levi-Civtita density is denoted by $\epsilon_{\alpha\beta}$ with $\epsilon_{01}=-\epsilon_{10}=1$ and $\epsilon_{00}=0=\epsilon_{11}$. From this we construct the 2-dimensional Levi-Civita tensor
\beq
\varepsilon_{\alpha\beta}=\sqrt{|g|}\epsilon_{\alpha\beta}
\eeq
and note the 2-dim identity
\beq
\varepsilon_{\alpha\beta}\varepsilon_{\mu\nu}=g_{\alpha\nu}g_{\beta\mu}-g_{\alpha\mu}g_{\beta\nu}
\eeq

\subsection{The Metric-Affine setup}
A generic $n-dim$  manifold endowed with a metric $g$ and an independent affine connection $\nabla$, denoted as ($\mathcal{M}$, g, $\nabla$) defines a metric-affine (non-Riemannian) geometry. On this extended  manifold, we define the curvature, torsion and non-metricity tensors according to \cite{schouten1954ricci}
\beq
R^{\mu}_{\;\;\;\nu\alpha\beta}:= 2\partial_{[\alpha}\Gamma^{\mu}_{\;\;\;|\nu|\beta]}+2\Gamma^{\mu}_{\;\;\;\rho[\alpha}\Gamma^{\rho}_{\;\;\;|\nu|\beta]} \label{R}
\eeq
\beq
S_{\mu\nu}^{\;\;\;\lambda}:=\Gamma^{\lambda}_{\;\;\;[\mu\nu]}
\eeq
\beq
Q_{\alpha\mu\nu}:=- \nabla_{\alpha}g_{\mu\nu}
\eeq
The departure of the affine connection $\Gamma^{\lambda}_{\;\;\;\mu\nu}$ from the Levi-Civita one defines the distortion tensor \cite{schouten1954ricci,hehl1995metric}
\begin{gather}
N^{\lambda}_{\;\;\;\;\mu\nu}:=\Gamma^{\lambda}_{\;\;\;\mu\nu}-\widetilde{\Gamma}^{\lambda}_{\;\;\;\mu\nu}=
\frac{1}{2}g^{\alpha\lambda}(Q_{\mu\nu\alpha}+Q_{\nu\alpha\mu}-Q_{\alpha\mu\nu}) -g^{\alpha\lambda}(S_{\alpha\mu\nu}+S_{\alpha\nu\mu}-S_{\mu\nu\alpha}) \label{N}
\end{gather}
where $\widetilde{\Gamma}^{\lambda}_{\;\;\;\mu\nu}$ is the usual Levi-Civita connection computed solely by the metric and its first derivatives. From the distortion, torsion and non-metricity can be retrieved according to (see for instance \cite{iosifidis2019metric})
\beq
S_{\mu\nu\alpha}=N_{\alpha[\mu\nu]}\;\;,\;\;\; Q_{\nu\alpha\mu}=2 N_{(\alpha\mu)\nu} \label{QNSN}
\eeq
Out of these we can construct the associated torsion and nonmetricity vectors  by taking the contractions:
\beq
S_{\mu}:=S_{\mu\lambda}^{\;\;\;\;\lambda} 
\eeq
and
\beq
Q_{\alpha}:=Q_{\alpha\mu\nu}g^{\mu\nu}\;,\;\; q_{\nu}=Q_{\alpha\mu\nu}g^{\alpha\mu}
\eeq
respectively.
As for the curvature tensor, we can construct the three contractions
\beq
R_{\nu\beta}:=R^{\mu}_{\;\;\nu\mu\beta}	
\;\;, \;\;
\hat{R}_{\alpha\beta}:=R^{\mu}_{\;\;\mu\alpha\beta}	
\;\;, \;\;
\breve{R}^{\mu}_{\;\;\beta}:=R^{\mu}_{\;\;\nu\alpha\beta}	g^{\nu\alpha}
\eeq
The first one is generalized Ricci tensor (which is not symmetric in general), the second is the so-called homothetic curvature  and the last one is known as the co-Ricci tensor. Observe that in constructing the Ricci scalar and homothetic curvature no metric is required, whereas in order to define the co-Ricci a metric should be given. The generalized scalar curvature is still uniquely defined, since
\beq
R:=g^{\mu\nu}R_{\mu\nu}=-g^{\mu\nu}\breve{R}_{\mu\nu} \;\; ,\;\;\; g^{\mu\nu}\hat{R}_{\mu\nu}=0
\eeq

Using eq. ($\ref{N}$) allows to split any quantity  into its Riemannian  (i.e. computed with respect to the Levi-Civita connection) and non-Riemannian contributions of torsion and nonmetricity. For instance inserting the  decomposition ($\ref{N}$) into the definition ($\ref{R}$) we obtain for the curvature tensor \footnote{Quantities with\;  $\widetilde{}$\;  will always denote Riemannian parts  unless otherwise stated.} 
\beq
{R^\mu}_{\nu \alpha \beta} = \widetilde{R}^\mu_{\phantom{\mu} \nu \alpha \beta} + 2 \widetilde{\nabla}_{[\alpha} {N^\mu}_{|\nu|\beta]} + 2 {N^\mu}_{\lambda|\alpha} {N^\lambda}_{|\nu|\beta]} \,, \label{decomp}
\eeq
Taking further contractions we can also find the post-Riemannian expansion of the  scalar curvature
\begin{gather}
R=\tilde{R}+ \frac{1}{4}Q_{\alpha\mu\nu}Q^{\alpha\mu\nu}-\frac{1}{2}Q_{\alpha\mu\nu}Q^{\mu\nu\alpha}    -\frac{1}{4}Q_{\mu}Q^{\mu}+\frac{1}{2}Q_{\mu}q^{\mu}+S_{\mu\nu\alpha}S^{\mu\nu\alpha}-2S_{\mu\nu\alpha}S^{\alpha\mu\nu}-4S_{\mu}S^{\mu} \nonumber \\ +2 Q_{\alpha\mu\nu}S^{\alpha\mu\nu}+2 S_{\mu}(q^{\mu}-Q^{\mu}) +\tilde{\nabla}_{\mu}(q^{\mu}-Q^{\mu}-4S^{\mu})
\end{gather}
\textcolor{black}{
Alternatively, in terms of the distortion, this expansion takes the much more compressed form
\beq
R=\tilde{R}+\widetilde{\nabla}_{\mu}(N^{\mu}{}_{\alpha\beta}g^{\alpha\beta}-N^{\alpha \mu}{}{}_{\alpha})+N_{\mu\alpha\beta}g^{\alpha\beta}N^{\gamma \mu}{}{}_{\gamma}-N_{\alpha\mu\nu}N^{\mu\nu\alpha}
\eeq
}

Finally, an object that we will be dealing with later on is the so-called Palatini tensor \cite{Hehl:1978zkk}, defined as the variation of the scalar curvature relative to the affine connection:
\begin{gather}
	P_{\lambda}^{\;\;\;\mu\nu}\textcolor{black}{:=\frac{\delta R}{\delta \Gamma^{\lambda}{}_{\mu\nu}}=}-\frac{\nabla_{\lambda}(\sqrt{-g}g^{\mu\nu})}{\sqrt{-g}}+\frac{\nabla_{\sigma}(\sqrt{-g}g^{\mu\sigma})\delta^{\nu}_{\lambda}}{\sqrt{-g}} 
	+2(S_{\lambda}g^{\mu\nu}-S^{\mu}\delta_{\lambda}^{\nu}+g^{\mu\sigma}S_{\sigma\lambda}^{\;\;\;\;\nu})  \nonumber \\
=\delta^{\nu}_{\lambda}\left( q^{\mu}-\frac{1}{2}Q^{\mu}-2 S^{\mu} \right) + g^{\mu\nu}\left( \frac{1}{2}Q_{\lambda}+2 S_{\lambda} \right)-( Q_{\lambda}^{\;\;\;\mu\nu}+2 S_{\lambda}^{\;\;\;\;\mu\nu}) .\label{Palatino}
	\end{gather}

\section{Torsion and non-Riemannian curvature in 2-dimensions}
In 2 dimensions, even for the generalized non-Riemannian background of Metric-Affine Geometry, many simplifications occur. The affine connection has now 8 independent components, torsion has 2 and non-metricity 6. Of course, the metric still has three components as in the Riemannian case but now the curvature has 4 independent components, the same number as the Ricci tensor. This suggests that, in principle, in this generalized geometric setting, one should be possible to determine the curvature from the Ricci tensor. As we explicitly prove in the following, this is indeed the case. Before proving this, we shall also review some known results for the  2-dimensional torsion, approached from a somewhat different perspective.

\subsection{2-dimensional Torsion}

The very fact that torsion has only 2 components implies that only its trace part is non-vanishing namely, it has the simple form
\beq
S_{\alpha\beta}{}^{\gamma}=2 S_{[\alpha}\delta^{\gamma}_{\beta]}\label{tor2d}
\eeq
with $S_{\alpha}:=S_{\alpha\kappa}{}{}^{\kappa}$ being the torsion vector introduced earlier.  Of course, even though this is intuitive by counting the degrees of freedom, a more formal proof is required. Let us prove this statement. 
Since in 2-dimensions any rank-3 totally antisymmetric tensor, or higher rank tensor with 3 indices antisymmetrized,  vanishes identically, we have
\beq
\delta^{\gamma}_{[\lambda}S_{\alpha\beta]}{}{}^{\kappa}=0.
\eeq
Expanding the latter and using also the antisymmetry property of the torsion tensor, 
it follows that
\beq
\delta^{\rho}_{\lambda}S_{\alpha\beta\kappa}+\delta^{\rho}_{\beta}S_{\lambda\alpha}{}{}^{\kappa}+\delta^{\rho}_{\alpha}S_{\beta\lambda}{}{}^{\kappa}=0
\eeq
Contracting in $\lambda=\kappa$ we readily obtain
\beq
S_{\alpha\beta}{}^{\gamma}=2 S_{[\alpha}\delta^{\gamma}_{\beta]}
\eeq
which is the stated result.
Note that no metric was used to derive this result; it holds true for a purely affine space as well, without the need of a metric. 
This torsion form implies that in 2-dim the contortion tensor assumes the simple form
\beq
K_{\alpha\mu\nu}:=S_{\mu\nu\alpha}-S_{\alpha\nu\mu}-S_{\alpha\mu\nu}=-4  S_{[\alpha}g_{\mu]\nu}
\eeq
Finally, it is trivial to show that for the above form of torsion, the following identities hold true
\begin{gather}
    \nabla_{k}S_{ij}{}{}^{k}=-2 \nabla_{[i}S_{j]} \;, \; \;\nabla_{[i}S_{j]}=\partial_{[i}S_{j]} \;, \;\; S_{ijk}S^{k}=0 \label{torrel}
\end{gather}
These relations will be found to be useful when we compute the various contractions of the non-Riemannian curvature. All of the above were well established standard results; let us now move on to obtain a new result for the  non-Riemannian curvature.

\section{2-dimensional Curvature}
We shall now focus on the curvature tensor of a 2-dimensional metric-affine space. In contrast to the Riemannian case, we now have the following.

\begin{theorem}

 In 2-dim space endowed with a general linear connection $\nabla$, the curvature tensor is completely determined by the generalized (asymmetric) Ricci tensor. Its coordinate expression reads
\beq
\boxed{R^{\lambda}{}_{\mu\beta\gamma}=2R_{\mu[\gamma}\delta^{\lambda}_{\beta]}}
\eeq
 
 \begin{proof}

Firstly, let us note that in deriving this expression no metric is required. Namely, the validity of the relation to be proved does not require the introduction of a metric but only that of a linear connection.\footnote{After all the notions of curvature and torsion are solely attached to the existence of a connection and do not rely on a metric. Given a connection one can compute its curvature and torsion without reference to any metric whatsoever.} Indeed,
starting from the dimension-dependent identity
\beq
R^{\kappa}{}_{\mu[\alpha\beta}\delta_{\gamma]}^{\lambda}=0
\eeq
expanding, and using also the antisymmetry of the curvature in its last indices it readily follows that
\beq
R^{\kappa}{}_{\mu\alpha\beta}\delta^{\lambda}_{\gamma}+R^{\kappa}{}_{\mu\gamma\alpha}\delta^{\lambda}_{\beta}+R^{\kappa}{}_{\mu\beta\gamma}=0
\eeq
Contracting the latter in $\kappa=\gamma$, we immediately get
\beq
R^{\lambda}{}_{\mu\alpha\beta}=-2R_{\mu[\alpha}\delta^{\lambda}_{\beta]} \label{Curvature2dim}
\eeq
as stated.
\end{proof}
\end{theorem}
Therefore we have:
\begin{cor} In a 2-dim non-Riemannian space, the curvature tensor is specified completely by the Ricci tensor.\footnote{One may anticipate this to be true even before plunge into calculations. Indeed, in n-dimensions the generic curvature tensor, given the only symmetry $R^{\lambda}{}_{\mu\alpha\beta}=-R^{\lambda}{}_{\mu\beta\alpha}$, has $\frac{n^{3}(n-1)}{2}$ independent components whereas the Ricci tensor $R_{\mu\nu}$ has $n^{2}$ independent components. The last number matches exactly the number of the curvature components only in 2-dimensions where both tensors have $4$ independent components. }
This construction does not rely on a metric. Furthermore, the vanishing of the scalar curvature does not necessarily imply that the Ricci and curvature tensors are zero.  
\end{cor}
\textbf{Remark.} Notice that no use of a metric was made in obtaining the last relation. The above result is also true for purely affine spaces, namely manifolds endowed with only a connection.
\newline

If a metric is given along with a connection (metric-affine space), then contracting (\ref{Curvature2dim}) with $g^{\mu\alpha}$, and bringing all indices down, it follows that
\beq
\breve{R}_{\kappa\beta}=-R g_{\kappa\beta}+R_{\kappa\beta} \label{RRhat}
\eeq
which relates the co-Ricci to the Ricci tensor and scalar curvature.  The antisymmetric part of the latter gives the relation $R_{[\nu\beta]}=\breve{R}_{[\nu\beta]}$. Expression (\ref{RRhat}) allows to express the curvature in the alternative form
\beq
R_{\alpha\beta\mu\nu}=\breve{R}_{\alpha\nu}g_{\mu\beta}-\breve{R}_{\alpha\mu}g_{\nu\beta} \label{Riemann2}
\eeq
In addition, (\ref{RRhat})  tells us that in contrast to the Riemannian case, in the 2-dim non-Riemannian geometry not all spaces are Einstein spaces. In terms of the Ricci tensor and scalar curvature, the curvature tensor reads
\beq
R_{\alpha\beta\mu\nu}=2R g_{\alpha[\mu}g_{\nu]\beta}-2 R_{\alpha[\mu}g_{\nu]\beta} \label{Curin2}
\eeq

\textbf{Remark.} It is important to stress, however, that even when we endow our space with a metric, in contradistinction with the Riemannian case, here the Ricci tensor  is not fully described only by the metric and the scalar curvature. Of course, this is easily seen from the very fact that $R_{\mu\nu}$ now also contains an antisymmetric part and therefore cannot be constructed only by the (symmetric) metric.

Therefore, given   equations (\ref{Curvature2dim}),(\ref{RRhat}) and (\ref{Curin2}) we have the following:
\begin{cor}
    If a 2-dimensional metric-affine space is either Ricci  flat ($R_{\mu\nu}=0$) or co-Ricci flat $(\breve{R}_{\mu\nu}=0)$ then it is completely flat, in the sense that $R^{\lambda}{}_{\mu\nu\alpha}=0$ in either case.
\end{cor}

\subsection{Ricci, co-Ricci and Homothetic Curvature}
By substituting the torsion relations (\ref{torrel}) into (\ref{decomp}) and taking the relevant contractions, it is then easy to see  that in the antisymmetric part of the Ricci tensor only non-metricity contributes since
\beq
R_{[ij]}=\frac{1}{2}\hat{R}_{ij} \label{R[]}
\eeq
for this form of torsion. Alternatively, this relation can also be  trivially derived by contracting (\ref{Curvature2dim}) in $\lambda=\mu$. Continuing, using again the fact that any totally antisymmetric rank-3 tensor (or any higher rank tensor antisymmetrized in three indices) identically vanishes in 2-dimensions, we have the identities
\beq
R_{[\alpha\beta\gamma]\lambda}=0
\eeq
Expanding the antisymmetrization and then
contracting  with $g^{\gamma\lambda}$ we find
\beq
\breve{R}_{[\alpha\beta]}=R_{[\alpha\beta]} \label{rR}
\eeq
then comparing the latter with (\ref{R[]}) it follows that all antisymmetric parts of the Ricci type tensors are proportional to one another.
Of course this is to be expected since in 2-dimensions any antisymmetric rank-2 tensor is a top-form and as such it is proportional to the antisymmetric Levi-Civita tensor $\epsilon_{\mu\nu}$. The proportionality factor is then fixed by employing the identity $\epsilon^{\mu\nu}\epsilon_{\mu\nu}=-2$ and we readily find
\beq
\breve{R}_{[\alpha\beta]}=R_{[\alpha\beta]}=\frac{1}{2}\hat{R}_{\alpha\beta}=\frac{1}{2}\partial_{[\alpha}Q_{\beta]}=-\frac{1}{2}(\epsilon^{\mu\nu}\partial_{\mu}Q_{\nu})\epsilon_{\alpha\beta} 
\eeq
Confirming that the antisymmetric parts of all Ricci-type tensors are proportional to one another.

\subsection{2-dimensional Scalar Curvature and Palatini tensor}

It is easy to check that in 2-dimensions the torsion contributions exactly cancel out and the Palatini tensor (\ref{Palatino}) acquires the form
\beq
P_{\alpha\mu\nu}=\frac{Q_{\alpha}}{2}g_{\mu\nu}-Q_{\alpha\mu\nu}+g_{\alpha\nu}\Big( q_{\mu}-\frac{Q_{\mu}}{2}\Big) \label{Palatini}
\eeq
This suggests that in 2-dim the scalar curvature is, probably up to total derivatives, independent of the torsion.\footnote{It is also interesting to observe that for Weyl-type non-metricity, i.e. when $Q_{\alpha\mu\nu}=\frac{1}{2}Q_{\alpha}g_{\mu\nu}$ the 2-dim Palatini tensor vanishes identically: $P_{\alpha\mu\nu}\equiv 0$. The explanation lies in the form of the scalar curvature, which in this case takes the form $R=\tilde{R}-\tilde{\nabla}_{\mu}\Big( \frac{1}{4}Q^{\mu}+4 S^{\mu}\Big)$ meaning that up to surface terms the scalar curvature is independent of both torsion and nonmetricity.} Indeed, given the form of torsion (\ref{tor2d}) it is a trivial task to show that in the post-Riemannian expansion of $R$ the quadratic torsion terms and mixed torsion-non-metricity terms exactly cancel out leaving us with
\beq
R=\tilde{R}+\frac{1}{4}Q_{\alpha\mu\nu}Q^{\alpha\mu\nu}-\frac{1}{2}Q_{\alpha\mu\nu}Q^{\mu\nu\alpha}    -\frac{1}{4}Q_{\mu}Q^{\mu}+\frac{1}{2}Q_{\mu}q^{\mu}+\tilde{\nabla}_{\mu}(q^{\mu}-Q^{\mu}-4 S^{\mu}) \label{Rpost}
\eeq
and consequently, up to total derivatives the scalar curvature is independent of torsion in 2-dim. Consequently the Einstein-Hilbert action at the linear level, does not include any torsion contribution. The last term in (\ref{Rpost}) that contains a torsion derivative will give contributions only if quadratic and higher order terms of the scalar curvature are considered or non-minimal couplings of the form $\Phi R$. In what follows, we shall investigate some of these possibilities in detail.

\section{A new symmetry of Scalar Curvature in 2-dim}
Under the displacement $\Gamma^{\lambda}{}_{\mu\nu} \rightarrow \Gamma^{\lambda}{}_{\mu\nu}+\Xi^{\lambda}{}_{\mu\nu}$
the curvature tensor transforms as \cite{iosifidis2019metric} 
\beq
R^{\mu}{}_{\nu\alpha\beta}\rightarrow R^{\mu}{}_{\nu\alpha\beta}+2 \nabla_{[\alpha}\Xi^{\mu}{}_{|\nu|\beta]}-2 S_{\alpha\beta}{}{}^{\lambda}\Xi^{\mu}{}_{\nu\lambda}+2 \Xi^{\mu}{}_{\lambda[\alpha}\Xi^{\lambda}{}_{|\nu|\beta]}
\eeq
Consequently, taking a double contraction (with $\delta^{\alpha}_{\mu}g^{\nu\beta}$) it follows that the scalar curvature transforms according to
\beq
R\rightarrow R+\nabla_{\alpha}(\Xi^{\alpha}_{(3)}-\Xi^{\alpha}_{(2)})+\Xi_{\mu}^{(2)}\Xi^{\mu}_{(3)}-\Xi_{\alpha\mu\nu}\Xi^{\mu\nu\alpha}+q^{\mu}\Xi_{\mu}^{(2)}-(Q_{\alpha\beta\gamma}+2 S_{\alpha\beta\gamma})\Xi^{\alpha\beta\gamma}
\eeq
where $\nabla$, $S$ and $Q$ are computed with respect to the initial connection and we have defined the three traces $\Xi_{\mu}^{(1)}:=\Xi_{\alpha\beta\mu}g^{\alpha\beta}$, $\Xi_{\mu}^{(2)}:=\Xi_{\alpha\mu\beta}g^{\alpha\beta}$ and $\Xi_{\mu}^{(1)}:=\Xi_{\mu\alpha\beta}g^{\alpha\beta}$.
Taking 
\beq
\Gamma^{\lambda}{}_{\mu\nu} \rightarrow \Gamma^{\lambda}{}_{\mu\nu}+\xi^{\lambda}g_{\mu\nu}-\delta^{\lambda}_{\nu}\xi_{\mu} \label{conntra2}
\eeq
it easily follows for arbitrary dimension n, that under ($\ref{conntra2}$), the density $\sqrt{-g}R$ transforms as follows: 
\beq
\sqrt{-g}R\rightarrow \sqrt{-g}R+2\partial_{\alpha}(\sqrt{-g}\xi^{\alpha})+(n-2)\sqrt{-g}\Big( 2\nabla_{\alpha}\xi^{\alpha}-(n-1)\xi^{\alpha}\xi^{\alpha}-\xi_{\alpha}q^{\alpha}\Big)
\eeq
therefore in  2-dimensions the last parenthesis drops out and
\beq
\sqrt{-g}R \rightarrow \sqrt{-g}R+2\partial_{\alpha}(\sqrt{-g}\xi^{\alpha})
\eeq
which means that (up to surface terms) the generalized Einstein Hilbert Lagrangian density remains invariant under (\ref{conntra2}). This implies that (see \cite{iosifidis2020linear}) the $\Gamma$-variation of $\sqrt{-g}R$ produces a tensor $P_{\lambda}{}^{\mu\nu}$ of which the trace in 1st and 3rd indices is the same with the trace in   2nd and 3rd, i.e. $P_{(2)}^{\alpha}=P^{\alpha}_{(3)}$. We will elaborate more on this result in the next sections.

When the latter observation is combined with the fact that the scalar curvature is projective invariant (in any dimension), namely invariant under the connection shifts
\beq
\Gamma^{\lambda}{}_{\mu\nu} \rightarrow \Gamma^{\lambda}{}_{\mu\nu}+\delta^{\lambda}_{\mu}\vartheta_{\nu} \label{projective}
\eeq
it immediately follows that for $n=2$ the scalar curvature enjoys a larger symmetry.
Indeed, given the fact that the connection transformations (\ref{conntra2}) and (\ref{projective}) are additive, we arrive to the following.

\begin{theorem}
    
 In 2-dimensions and up to total derivative terms, the scalar curvature $R$ is invariant under connection transformations of the form
\beq
\Gamma^{\lambda}{}_{\mu\nu} \rightarrow \Gamma^{\lambda}{}_{\mu\nu}+\delta^{\lambda}_{\mu}\vartheta_{\nu}+\xi^{\lambda}g_{\mu\nu}-\delta^{\lambda}_{\nu}\xi_{\mu} \label{conntra3}
\eeq
where $\vartheta_{\mu}$ is an arbitrary 1-form and $\xi^{\lambda} $ an arbitrary vector field. This enlarged symmetry is a specific feature of 2-dimensions.\footnote{For $n\geq 3$ the only symmetry for the scalar curvature is the projective symmetry, namely invariance under (\ref{projective}) alone.}
\end{theorem}

\section{Classification of generalized Einstein spaces in 2-dim}

\subsection{2-dimensional Riemann-Cartan Geometry}
It is interesting to see what changes once we restrict the non-Riemannian background by demanding also the metricity condition, i.e. imposing vanishing non-metricity. We are then in the realm of Riemann-Cartan geometry \cite{Cartan1922,Cartan1923,Cartan1924,Cartan1925}. In this case, it follows immediately that the full distortion is completely specified by the torsion vector (along with the metric) and reads
\beq
N^{\lambda}{}_{\mu\nu}=-2 (S^{\lambda}g_{\mu\nu}-\delta^{\lambda}_{\nu}S_{\mu}) \label{Ntor}
\eeq
With this restriction we have the following.

\begin{theorem}
Every 2-dimensional manifold with a metric compatible but torsionful connection, is an Einstein space in the sense that
\beq
R_{\mu\nu}=\frac{R}{2}g_{\mu\nu}
\eeq
with the scalar curvature  given by
\beq
R=\tilde{R}-4 \tilde{\nabla}_{\mu}S^{\mu}
\eeq
where $\tilde{R}$ is the Riemannian scalar curvature (Riemannian Ricci scalar), $\tilde{\nabla}_{\mu}$ the Levi-Civita covariant derivative, and $S^{\mu}$ the torsion vector. In other words, every 2-dimensional Riemman-Cartan manifold is an Einstein space.
\end{theorem}

\begin{proof}
Starting from the definition of the curvature tensor and substituting in the form of the distortion (\ref{Ntor})  it readily follows that
\beq
R^{\mu}{}_{\nu\alpha\beta}=\tilde{R}^{\mu}{}_{\nu\alpha\beta}+4 \delta^{\mu}_{[\beta}(\tilde{\nabla}_{\alpha]}-2 S_{\alpha]})S_{\nu}-4 g_{\nu[\beta}(\tilde{\nabla}_{\alpha]}-2 S_{\alpha]}) S^{\mu}-8 \delta^{\mu}_{[\alpha}g_{\beta]\nu}S_{\gamma}S^{\gamma}
\eeq
Contracting  we obtain the associated Ricci tensor\footnote{In general dimensions for vectorial torsion this reads
\beq
R_{\mu\nu}=\tilde{R}_{\mu\nu}-\frac{2}{(d-1)}g_{\mu\nu}\tilde{\nabla}_{\alpha}S^{\alpha}-\frac{2(d-2)}{(d-1)}\tilde{\nabla}_{\nu}S_{\mu}+\frac{4(d-2)}{(d-1)^{2}}(S_{\mu}S_{\nu}-S^{2}g_{\mu\nu})
\eeq
But of course, it goes without saying,  that in higher dimensions there are more modes to the torsion tensor apart from its trace.
}
\beq
R_{\mu\nu}=\tilde{R}_{\mu\nu}-2 g_{\mu\nu}\tilde{\nabla}_{\alpha}S^{\alpha}\label{Rtor}
\eeq
Then, given that 
\beq
\tilde{R}_{\mu\nu}=\frac{\tilde{R}}{2}g_{\mu\nu}
\eeq
for the Riemannian part in 2 dimensions,
substitution of the latter into (\ref{Rtor}) gives the stated result.
\end{proof}

\subsection{2-dimensional Riemann-Cartan-Weyl Geometry}

Let us now consider a 2-dimensional Riemann-Cartan-Weyl geometry. Traditionally, a Riemann-Cartan-Weyl geometry is characterised by an arbitrary torsion but with a restricted Weyl-type non-metricity, viz.\footnote{For general dimensions we have the form $Q_{\alpha\mu\nu}=\frac{Q_{\alpha}}{n}g_{\mu\nu}$.}
\beq
Q_{\alpha\mu\nu}=\frac{Q_{\alpha}}{2}g_{\mu\nu}
\eeq
for such geometries. Therefore this is the simplest non-trivial extension of the previous Riemann-Cartan case.

Now, as we have already seen in 2-dimensions torsion is purely vectorial. Combining the above two results it follows that in 2-dimensional Weyl-Cartan spaces, the distortion tensor takes the form
\beq
N^{\lambda}{}_{\mu\nu}=\delta^{\lambda}_{\mu}\frac{1}{4}Q_{\nu}+\delta^{\lambda}_{\nu}w_{\mu}-w^{\lambda}g_{\mu\nu} \label{distorW}
\eeq
where $w_{\mu}=Q_{\mu}/4+2 S_{\mu}$. 
Then it follows that:
\begin{theorem}
    Every 2-dimensional Weyl-Cartan-Riemann space is a generalized Einstein space in the sense that 
    \beq
R_{(\mu\nu)}=\frac{R}{2}g_{\mu\nu}
    \eeq  
\end{theorem}
is identically valid.

\begin{proof}

For the distortion (\ref{distorW}) we readily compute the associated Ricci tensor
\beq
R_{\mu\nu}=\tilde{R}_{\mu\nu}+\frac{1}{2}\partial_{[\mu}Q_{\nu]}-(\nabla_{\alpha}w^{\alpha})g_{\mu\nu}
\eeq
Note that $R_{[\mu\nu]}=\frac{1}{2}\partial_{[\mu}Q_{\nu]}=\frac{1}{2} \hat{R}_{\mu\nu}$ as it should in 2-dimensions, in agreement with the results of the previous section.
The associated scalar curvature to the above Ricci tensor, reads
\beq
R=\tilde{R}-2\tilde{\nabla}_{\alpha}w^{\alpha}
\eeq
Then given the fact that $\tilde{R}_{\mu\nu}=\frac{\tilde{R}}{2}g_{\mu\nu}$ it immediately follows that
\beq
R_{(\mu\nu)}=\frac{R}{2}g_{\mu\nu}
\eeq
    \end{proof}
\begin{cor}
    Any 2-dimensional Weyl-Cartan geometry is a generalized Einstein space. 
\end{cor}

 To gain more intuition on this result, for generic nonmetricity we may separate out the Weyl part and decompose the nonmetricity according to
 \beq
Q_{\alpha\mu\nu}=\frac{Q_{\alpha}}{2}+\Omega_{\alpha\mu\nu}
 \eeq
with $\Omega_{\alpha\mu\nu}g^{\mu\nu}=0$. Then, a simple substitution of the latter into the decomposition (\ref{Rpost}) yields
\beq
R=\tilde{R}+\frac{1}{2}\Omega_{\alpha\mu\nu}\Omega^{\alpha\mu\nu}-\frac{1}{2}\Omega_{\alpha\mu\nu}\Omega^{\mu\nu\alpha}+\tilde{\nabla}_{\mu}(q^{\mu}-Q^{\mu}-4 S^{\mu})
\eeq
meaning that the Weyl part contributes only a total derivative term to the scalar curvature and the rest of the quadratic Weyl contributions disappear.  Consequently, the variation of the scalar curvature is completely independent of the Weyl vector (as is also from the torsion vector) and takes the form 
\beq
P_{\alpha\mu\nu}=-\Omega_{\alpha\mu\nu}+g_{\alpha\nu}\Omega_{\beta\gamma\mu}g^{\beta\gamma}
\eeq
We see therefore that both the scalar curvature and its variation (Palatini tensor) depend only on the non-Weyl part of nonmetricity. It is quite interesting that the same part is responsible for the non-topological character of the generalized Gauss-Bonnet term \cite{BabourovaFrolov1997}.

\section{Setting up Metric-Affine Theories in 2-dim}

For general dimensions, out of the curvature tensor and its contractions one can form 16 quadratic parity even invariants \cite{Vassiliev:2003dk} and the full curvature-quadratic action reads
\begin{gather}
S=\frac{1}{2 \kappa} \int d^{n}x \sqrt{-g}\Big[ R +a_{1}R^{2}+a_{2}R_{\mu\nu}R^{\mu\nu}+ a_{3}R_{\mu\nu}R^{\nu\mu}+a_{4}\breve{R}_{\mu\nu}\breve{R}^{\mu\nu}+ a_{5}\breve{R}_{\mu\nu}\breve{R}^{\nu\mu}+a_{6}\hat{R}_{\mu\nu}\hat{R}^{\mu\nu} \nonumber \\
a_{7}R_{\mu\nu}\breve{R}^{\mu\nu}+a_{8} R_{\mu\nu}\breve{R}^{\nu\mu}+a_{9}R_{\mu\nu}\hat{R}^{\mu\nu}+a_{10}\breve{R}_{\mu\nu}\hat{R}^{\mu\nu}+a_{11}R_{\mu\nu\alpha\beta}R^{\mu\nu\alpha\beta} \nonumber \\
+a_{12}R_{\mu\nu\alpha\beta}R^{\nu\alpha\mu\beta}+a_{13}R_{\mu\nu\alpha\beta}R^{\nu\mu\alpha\beta}+a_{14}R_{\mu\nu\alpha\beta}R^{\alpha\nu\mu\beta}+a_{15}R_{\mu\nu\alpha\beta}R^{\mu\alpha\nu\beta}+a_{16}R_{\mu\nu\alpha\beta}R^{\alpha\beta\mu\nu} \Big] \label{actionR2}
\end{gather}
In 2-dimensions, however, this number is sufficiently reduced as one may anticipate. Indeed, the very form of the curvature tensor (\ref{Curin2}) immediately implies that all quadratic curvature invariants would be linear combinations of $R^{2}$, $R_{\mu\nu}R^{\mu\nu}$ and $R_{\mu\nu}R^{\nu\mu}$. 
Therefore, only the first 3 quadratic terms in the first line of (\ref{actionR2}) are relevant in 2-dimensions, and the rest of the terms will just shift the coefficients $a_{1}, a_{2}$ and $a_{3}$. Then, after renaming the aforementioned parameters, we may then safely set the rest to zero, i.e. $a_{i}=0$, $i=4,5,...,16$. Therefore, in 2-dimensions the most general action that is quadratic in curvature\footnote{Of course along with the standard Einstein-Hilbert term that is linear in curvature.} is given by
\beq
 \mathcal{L}^{(2)}=R +a_{1}R^{2}+a_{2}R_{\mu\nu}R^{\mu\nu}+ a_{3}R_{\mu\nu}R^{\nu\mu}
\eeq
namely, it is specified by three parameters only.  
Consequently, in 2-dimensions the most general Metric-Affine up to second-order  curvature invariants, reads
\beq
S^{(2)}[g,\Gamma]=\frac{1}{2 \kappa} \int d^{n}x \sqrt{-g}\Big[ R +a_{1}R^{2}+a_{2}R_{\mu\nu}R^{\mu\nu}+ a_{3}R_{\mu\nu}R^{\nu\mu} \Big]
\eeq
It should be noted that there will be an undetermined scalar mode in this theory. This is not special in 2-dim, any MAG Theory, in arbitrary dimensions,  constructed solely by the curvature tensor (and its various contractions) has this property \cite{schouten1954ricci}. This is easily seen from the fact that in any dimensions the curvature tensor is invariant under special projective transformations
\beq
\Gamma^{\lambda}{}_{\mu\nu}\mapsto \Gamma^{\lambda}{}_{\mu\nu}+\delta^{\lambda}_{\mu}\partial_{\nu}\vartheta
\eeq
with $\vartheta$ being an arbitrary scalar function. This invariance then means that there will be an undetermined scalar mode which, however, can be fixed by gauge choice.

Varying with respect to the metric and  the independent affine connection we derive the field equations
\begin{gather}
-\frac{1}{2}g_{\mu\nu}\Big( R +a_{1}R^{2}+a_{2}R_{\mu\nu}R^{\mu\nu}+ a_{3}R_{\mu\nu}R^{\nu\mu} \Big)+\Big(1+2 a_{1}R\Big)R_{(\mu\nu)} \nonumber \\
+a_{2}\Big( R_{\mu}{}^{\alpha}R_{\nu\alpha}+R_{\alpha\mu}R^{\alpha}{}_{\nu}\Big)+2 a_{3} R^{\alpha}{}_{(\mu}R_{\nu)\alpha}=0 \label{mfe1}
\end{gather}
and 
\beq
(-\nabla_{\alpha}+2 S_{\alpha})(\sqrt{-g}\Omega_{\lambda}{}^{\mu\alpha\nu})-\sqrt{-g}\Omega_{\lambda}{}^{\mu\beta\gamma}S_{\gamma\delta}{}{}^{\nu}=0
\eeq
respectively,  
where
\beq
\Omega_{\lambda}{}^{\mu\alpha\nu}=\frac{\partial(R+\mathcal{L}^{(2)})}{\partial R^{\lambda}{}_{\mu\alpha\nu}}=(1+2 a_{1}R)\delta_{\lambda}^{[\alpha}g^{\nu]\mu}+2 a_{2}R^{\mu[\nu}\delta^{\alpha]}_{\lambda}+ 2 a_{3}\delta^{[\alpha}_{\lambda}R^{\nu]\mu}
\eeq
Let us firstly focus on the connection field equations.
In 2-dimensions and given the form of the torsion tensor (\ref{tor2d}) , the above connection field equations simplify considerably and take the form
\beq
\nabla_{\alpha}(\sqrt{-g}\Omega_{\lambda}{}^{\mu\alpha\nu})=0
\eeq
as can be trivially checked. We see that the terms explicit in torsion have canceled out. Let us observe that $\Omega_{\lambda}{}^{\mu\alpha\nu}$ can be written as
\beq
\Omega_{\lambda}{}^{\mu\alpha\nu}=\delta_{\lambda}^{[\alpha}H^{\nu]\mu}
\eeq
where 
\beq
H^{\nu\mu}=(1+2 a_{1}R)g^{\nu\mu}+2 a_{2}R^{\mu\nu}+2 a_{3}R^{\nu\mu}
\eeq
Using this, the above connection field equations become
\beq
\nabla_{\lambda}(\sqrt{-g}H^{\nu\mu})-\delta_{\lambda}^{\nu}\nabla_{\alpha}(\sqrt{-g}H^{\alpha\mu})=0 \label{Homega}
\eeq
Contracting in $\nu=\lambda$ it follows that
\beq
\nabla_{\alpha}(\sqrt{-g}H^{\alpha\mu})=0
\eeq
which when substituted back to (\ref{Homega}) results in
\beq
\nabla_{\lambda}(\sqrt{-g}H^{\nu\mu}) =0
\eeq
implying that the tensor density $\sqrt{-g}H^{\nu\mu}$ is covariantly constant on-shell.

Now, let us turn our attention to the metric field equations. Taking the trace of (\ref{mfe1}) we obtain
\beq
a_{1}R^{2}+a_{2}R_{\mu\nu}R^{\mu\nu}+ a_{3}R_{\mu\nu}R^{\nu\mu} =0
\eeq
This is an algebraic equation for the traces of $R_{\mu\nu}$. Indeed, if we denote $R_{\mu\nu}$ by the matrix $\textbf{A}$ the above can be written as 
\beq
a_{1}\Big(Tr(A)\Big)^{2}+a_{2}Tr(A A^{T})+a_{3}Tr(A^{2})=0
\eeq
and may be   used to eliminate one of them. Substituting this back to the metric field equations (\ref{mfe1}), we arrive at the simplified form
\beq
-\frac{1}{2}g_{\mu\nu}R+\Big(1+2 a_{1}R\Big)R_{(\mu\nu)} \nonumber \\
+a_{2}\Big( R_{\mu}{}^{\alpha}R_{\nu\alpha}+R_{\alpha\mu}R^{\alpha}{}_{\nu}\Big)+2 a_{3} R^{\alpha}{}_{(\mu}R_{\nu)\alpha}=0 
\eeq
This is fairly general. Let us consider some special cases.

\subsection{The case $a_{2}=a_{3}=0$.}

We shall first consider the case where only the scalar curvature is included, that is, we set $a_{2}=a_{3}=0$. Then the metric and connection field equations reduce to
\beq
(1+2 a_{1}R)R_{(\mu\nu)}-\frac{1}{2}(R+a_{1}R^{2})g_{\mu\nu}=0 \label{go}
\eeq
\beq
\nabla_{\lambda}\Big( \sqrt{-g}(1+2 a_{1}R)g^{\mu\nu}\Big)=0
\eeq
respectively. Taking the trace of the above metric field equations, for $a_{1}\neq 0$ we deduce
\beq
R=0
\eeq
which, when substituted back to (\ref{go}), implies that
\beq
R_{(\mn)}=0
\eeq
Given that $R=0$, the connection field equations simplify to
\beq
\nabla_{\lambda}( \sqrt{-g}g^{\mu\nu})=0
\eeq
It is important to note that in 2-dimensions the latter does not imply a vanishing non-metricity but rather that
\beq
Q_{\alpha\mu\nu}=\frac{Q_{\alpha}}{2}g_{\mn}
\eeq
namely we have non-metricity of the Weyl-type. Since in 2-dim torsion is always of a purely vectorial nature we have 2 vectors completely specifying the non-Riemannian degrees of freedom. However, notice that since the action we started with consists only of the scalar curvature, enjoys the enlarged symmetry (\ref{conntra3}). Therefore we can choose $\vartheta_{\mu}$ and $\xi_{\mu}$ appropriately so as to obtain vanishing torsion and non-metricity vectors. In these particular gauges we have
\beq
S_{\mu}=0 \;\;, \;\; Q_{\mu}=0
\eeq
and subsequently the full torsion and non-metricity vanish, viz.
\beq
S_{\alpha\mn}=0\;\;, \;\; Q_{\lambda\mn}=0
\eeq
Then, given the post-Riemannian decomposition (\ref{Rpost}) of $R$ it follows that
\beq
\tilde{R}=0
\eeq
and consequently $\tilde{R}_{\mn}=0$. Therefore there is no dynamics for this Theory. The dynamics arises once we include the rest of the Ricci tensor pieces or by supplementing the action with quadratic torsion and non-metricity invariants which manifestly brake the symmetry of the action under connection transformations of the form (\ref{conntra3}).

\section{Full Quadratic Theory/Including Quadratic Torsion and Non-metricity terms}

Theories built only out of the curvature are by themselves interesting however they are restrictive in some sense. Indeed, in the non-Riemannian setup of MAG, along with the curvature the space possesses also the geometric features of torsion and non-metricity. Additionally, as we saw earlier, Theories that are built solely from the curvature tensor have the special projective symmetry which implies the existence of an undetermined scalar mode. Therefore we must also supplement our gravitational action with invariants built out of torsion and non-metricity. In general (i.e. for any dimension) there exist 11 parity even quadratic invariants in torsion and non-metricity. Of course in 2-dimensions this number is reduced substantially.\footnote{For instance the 3 quadratic pure torsion scalars degenerate all to $S_{\mu}S^{\mu}$ given the purely vectorial form of torsion in 2-dimensions.}To maintain simplicity and also make contact with the existing literature it suffices to consider the following  two quadratic terms
\beq
S_{\mu}S^{\mu}\;\;, \;\; Q_{\mu}Q^{\mu}
\eeq
Let us note that adding more quadratic invariants to the action apart from the above two does not change the qualitative behavior of the Theory and  only shifts  the various coefficients by introducing more parameters. So, our extended  quadratic Theory is now given by
\beq
S^{(2)}[g,\Gamma]=\frac{1}{2 \kappa} \int d^{n}x \sqrt{-g}\Big[ R +a_{1}R^{2}+a_{2}R_{\mu\nu}R^{\mu\nu}+ a_{3}R_{\mu\nu}R^{\nu\mu}+b_{1}S_{\mu}S^{\mu}+b_{2}Q_{\mu}Q^{\mu} \Big] \label{S}
\eeq
This is the starting point for obtaining non-trivial dynamics.

\subsection{Special Cases}
Let us now focus on the special case $a_{2}=a_{3}=0$ which was studied in \cite{Obukhov:2003tz}. Here we shall verify the conclusions found there but in a holonomic frame. In this special case the connection field equations read
\beq
\nabla_{\lambda}(\sqrt{-g}H^{\nu\mu})-\delta_{\lambda}^{\nu}\nabla_{\alpha}(\sqrt{-g}H^{\alpha\mu})+\sqrt{-g}\Big( 2 b_{1}S^{[\mu}\delta^{\nu]}_{\lambda}+ 4 b_{2} \delta^{\mu}_{\lambda}Q^{\nu}\Big)=0 \label{ConFESQ}
\eeq
where now $H^{\nu\mu}=(1+2 a_{1}R)g^{\nu\mu}$.
Tracing in $\lambda=\nu$ it follows that
\beq
\nabla_{\alpha}(\sqrt{-g}H^{\alpha\mu})=\sqrt{-g}(b_{1}S^{\mu}+4 b_{2}Q^{\mu})
\eeq
which when substituting back to the above brings it to the form
\beq
\frac{1}{\sqrt{-g}}\nabla_{\lambda}(\sqrt{-g}H^{\nu\mu})-b_{1}\delta_{\lambda}^{\mu}S^{\nu}+8 b_{2}\delta_{\lambda}^{[\mu}Q^{\nu]}=0\label{conreduc}
\eeq
Taking the antisymmetric part of the latter in $\mu,\nu$ fixes
\beq
b_{1}S_{\mu}=8 b_{2}Q_{\mu} \label{SQ}
\eeq
Substituting this into (\ref{conreduc}) and using the Leibniz rule for the covariant derivative, remembering also the definition of non-metricity we find
\beq
-\frac{Q_{\lambda}}{2}(1+2 a_{1}R)g^{\mu\nu}+2 a_{1}(\partial_{\lambda}R)g^{\mu\nu}+(1+2 a_{1}R)Q_{\lambda}{}^{\mu\nu}=8 b_{2}\delta^{(\mu}_{\lambda}Q^{\nu)} \label{NM}
\eeq
Contracting this one time with $g_{\mu\nu}$ and another in $\nu=\lambda$ we obtain
\beq
Q_{\mu}=\frac{a_{1}}{2 b_{2}}\partial_{\mu}R \label{QR}
\eeq
and 
\beq
q_{\mu}=\Big[ \frac{1}{2}+\frac{8 b_{2}}{1+2 a_{1}R}\Big]Q_{\mu}=\Big[ \frac{1}{2}+\frac{8 b_{2}}{1+2 a_{1}R}\Big]\frac{a_{1}}{2 b_{2}}\partial_{\mu}R=\partial_{\mu}\Big( \frac{a_{1}}{4 b_{2}}R+2 \ln{(1+2 a_{1}R)}\Big)
\eeq
In addition, combining (\ref{SQ}) with (\ref{QR}) it follows that
\beq
S_{\mu}=\frac{4 a_{1}}{b_{1}}\partial_{\mu}R
\eeq
Consequently, all three torsion and non-metricity vectors are proportional to one another and powered from the scalar curvature. Note that all three vectors are exact. Using the above forms of these vectors we can  exactly solve  (\ref{NM}) for non-metricity 
\beq
Q_{\alpha\mu\nu}=a_{1}\left[ \frac{1}{4 b_{2}}-\frac{2}{1+2 a_{1}R}\right](\partial_{\alpha}R)g_{\mu\nu}+\frac{4 a_{1}}{1+2 a_{1}R}g_{\alpha(\mu}\partial_{\nu)}R \label{Qdyn}
\eeq
given also the pure vectorial type of torsion in 2-dim the full torsion reads
\beq
S_{\mu\nu\alpha}=\frac{8 a_{1}}{b_{1}}g_{\alpha[\nu}\partial_{\mu]}R \label{Sdyn}
\eeq
Therefore we see that the entire non-metricity tensors are given in terms of the derivative of the scalar curvature. Let us now  find the evolution equation that the latter satisfies. To this end, we now turn our attention to the metric field equations, which in this case take the form
\begin{gather}
    (1+2 a_{1}R)R_{(\mu\nu)}-\frac{1}{2}(R+a_{1}R^{2})g_{\mu\nu}+b_{1}\Big( S_{\mu}S_{\nu}-\frac{1}{2}S_{\alpha}S^{\alpha}g_{\mu\nu} \Big)\nonumber \\+b_{2}\Big( Q_{\mu}Q_{\nu}-\frac{1}{2}Q_{\alpha}Q^{\alpha}g_{\mu\nu} \Big)-2 b_{2}g_{\mu\nu}\frac{1}{\sqrt{-g}}\partial_{\alpha}(\sqrt{-g}Q^{\alpha})=0
\end{gather}
Taking the trace and using (\ref{QR}) it follows that
\beq
R^{2}- \frac{2}{\sqrt{-g}}\partial_{\alpha}(\sqrt{-g}\partial^{\alpha}R)=0
\eeq
This is a dynamical Klein-Gordon-like equation for the scalar curvature, which given also the relations (\ref{Qdyn}) and (\ref{Sdyn})  gives dynamics to torsion and nonmetricity as well. Therefore we have non-trivial dynamics!
It is worth stressing that, quite remarkably, all parameters have disappeared from the final equation.

\subsection{Case $\alpha_{3}=-a_{2}$}
Let us now extend our previous considerations and allow for non-vanishing $a_{2}$ and $a_{3}$ that obey the relation $a_{3}=-a_{2}$. This choice picks out the homothetic curvature part from the quadratic Ricci terms, namely the tensor $H^{\nu\mu}$ now becomes
\beq
H^{\nu\mu}=(1+2 a_{1}R)g^{\mu\nu}+a_{2}\hat{R}^{\mu\nu}\label{Hsq}
\eeq
Then the connection field equations our formalistically the same with \eqref{ConFESQ} but for this form of $H^{\nu\mu}$. Then similarly to the previous  we first take the trace of the connection field equations in $\lambda=\nu$ which gives
\beq
\nabla_{\alpha}(\sqrt{-g}H^{\alpha\mu})=\sqrt{-g}(b_{1}S^{\mu}+4 b_{2}Q^{\mu})
\eeq
which again when substituted back to \eqref{ConFESQ} leaves us with
\beq
\frac{1}{\sqrt{-g}}\nabla_{\lambda}(\sqrt{-g}H^{\nu\mu})-b_{1}\delta_{\lambda}^{\mu}S^{\nu}+8 b_{2}\delta_{\lambda}^{[\mu}Q^{\nu]}=0 \label{Cfehere}
\eeq
Taking the symmetric part $(\nu\mu)$ of the latter, expanding the covariant derivative, and using \eqref{Hsq} we obtain
\beq
-\frac{Q_{\lambda}}{2}(1+2 a_{1}R)g^{\mu\nu}+2 a_{1}(\partial_{\lambda}R)g^{\mu\nu}+(1+2 a_{1}R)Q_{\lambda}{}^{\mu\nu}=b_{1}\delta^{(\mu}_{\lambda}S^{\nu)} \label{QQR}
\eeq
Contracting with $g_{\mu\nu}$ it follows that
\beq
S_{\lambda}=\frac{4 a_{1}}{b_{1}}\partial_{\lambda}R
\eeq
Using this, if we now contract (\ref{QQR}) in $\lambda=\nu$ we get
\beq
(1+2 a_{1}R)\Big( q^{\mu}-\frac{1}{2}Q^{\mu}\Big)=b_{1}S^{\mu}=4 a_{1}\partial^{\mu}R
\eeq
Furthermore, if we now take the antisymmetric part $[\mu\nu]$ of the connection field equations \eqref{Cfehere} and subsequently contract in $\lambda=\mu$ we obtain
\beq
a_{2}\frac{1}{\sqrt{-g}}\partial_{\mu}(\sqrt{-g}\hat{R}^{\mu\nu})=2 a_{1}\partial^{\nu}R-4 b_{2}Q^{\nu} \label{firRQ}
\eeq
Since $\hat{R}^{\mu\nu}$ is antisymmetric in its indices, if we act with another derivative, it follows that
\beq
2 b_{2}\frac{1}{\sqrt{-g}}\partial_{\mu}(\sqrt{-g}Q^{\mu})=a_{1}\Box_{g}R \label{RQrel}
\eeq

Let us now focus on the metric field equations which read
\begin{gather}
    -\frac{1}{2}\Big( R+a_{1} R^{2}+\frac{a_{2}}{2}\hat{R}_{\alpha\beta}\hat{R}^{\alpha\beta} \Big)g_{\mu\nu}+(1+2 a_{1}R) R_{(\mn)}+a_{2}\hat{R}_{\mu\alpha}\hat{R}_{\nu\beta}g^{\ab} \nonumber \\b_{1}\Big( S_{\mu}S_{\nu}-\frac{1}{2}S_{\alpha}S^{\alpha}g_{\mu\nu} \Big)\nonumber +b_{2}\Big( Q_{\mu}Q_{\nu}-\frac{1}{2}Q_{\alpha}Q^{\alpha}g_{\mu\nu} \Big)-2 b_{2}g_{\mu\nu}\frac{1}{\sqrt{-g}}\partial_{\alpha}(\sqrt{-g}Q^{\alpha})=0
\end{gather}
Taking the trace we obtain
\beq
a_{1}R^{2}+\frac{a_{2}}{2}\hat{R}_{\ab}\hat{R}^{\ab}=4b_{2}\frac{1}{\sqrt{-g}}\partial_{\alpha}(\sqrt{-g}Q^{\alpha})
\eeq
which in view of (\ref{RQrel}) can be written as
\beq
2 \Box_{g}R=R^{2}+\frac{a_{2}}{2 a_{1}}\hat{R}_{\ab}\hat{R}^{\ab}\label{secRQ}
\eeq
We see therefore that the collection of (\ref{firRQ}) and (\ref{secRQ}) describes the propagation of a scalar ($R$) and a vector ($Q^{\mu}$) coupled with each other. Given the solutions of $R$ and $Q^{\mu}$ one can then compute the full torsion and non-metricity tensors.
This result is quite interesting; however, it goes beyond the purpose of this study  to further investigate the solutions of this Theory.  The purpose was to show how the inclusion of torsion and nonmetricity, provides non-trivial dynamics for 2-dimensional gravity.

\section{Conclusions}

We have studied the structure of 2-dimensional Metric-Affine Geometries consisting of curvature, torsion and non-metricity. We have explicitly proved that in such geometries, in contrast to the Riemannian case, the Riemann tensor is not completely determined by the scalar curvature, but rather by the Ricci tensor which is now asymmetric. Furthermore, we have found a new symmetry for the generalized scalar curvature in 2-dimensions, the invariance under the connection transformations (\ref{conntra3}).

Moving on to the physical applications, we used the above results to study 2-dimensional Metric-Affine Theories in vacuum. Considering quadratic Yang-Mills type Theories in curvature, torsion and nonmetricity, we explicitly showed how the extended geometry offers the possibility to have propagating degrees of freedom.  In particular, we presented two examples where torsion and nonmetricity are propagating in 2-dimensions.

It would be interesting to study the integrability and black hole solutions of our general theory (\ref{S}), compared to the Poincare case \cite{Mielke1993} . Needless to say, the theoretical framework we developed here can be applied to study the whole plethora of 2-dim gravity applications such as holography  \cite{AlmheiriPolchinski2015,MaldacenaStanfordYang2016,
MertensTuriaci2023},
string theory 
\cite{DiFrancesco1995,DistlerKawai1989,David1988},
cosmology
\cite{Bhattacharya2020,MazzitelliRusso1993,Bose:1997qv} and so on.

\section{Acknowledgments}
 This work was supported by the  Istituto Nazionale di Fisica Nucleare (INFN), Sezioni  di Napoli,  {\it Iniziative Specifiche} SKY.

\bibliographystyle{unsrt}
	\bibliography{ref}

\end{document}